\documentclass[a4paper,11pt]{article}
\pdfoutput=1 

\usepackage{jheppub} 

\usepackage[T1]{fontenc} 
\usepackage{graphicx}
\usepackage{epsfig}  
\usepackage{epsf}    
\usepackage{dcolumn}
\usepackage{bm}
\usepackage{dcolumn}
\usepackage{textcomp}
\usepackage{float}
\usepackage{subfig}
\usepackage[]{hyperref}
  \hypersetup{
  unicode=false,          
  pdftoolbar=true,        
  pdfmenubar=true,        
  pdffitwindow=true,     
  pdfstartview={FitH},    
  pdfnewwindow=true,      
  pdfcreator={RevTeX},
  colorlinks=true,       
  linkcolor=red,          
  citecolor=blue,        
  urlcolor=blue,           
  }
\usepackage{hypcap}

\title{New physics solutions for $R_D$ and $R_{D^*}$}
\author[a]{Ashutosh Kumar Alok}
\author[b]{Dinesh Kumar}
\author[c]{Jacky Kumar}
\author[d]{Suman Kumbhakar}
\author[d]{S. Uma Sankar}

\affiliation[a]{Indian Institute of Technology Jodhpur, Jodhpur 342011, India}
\affiliation[b]{Department of Physics, University of Rajasthan, Jaipur 302004, India}
\affiliation[c]{Department of High Energy Physics, Tata Institute of Fundamental Research, 400005, Mumbai, India}
\affiliation[d]{Department of Physics, Indian Institute of Technology Bombay,
Mumbai 400076, India}
\emailAdd{akalok@iitj.ac.in}
\emailAdd{dinesh@uniraj.ac.in}
\emailAdd{jkumar@iisermohali.ac.in}
\emailAdd{suman@phy.iitb.ac.in}
\emailAdd{uma@phy.iitb.ac.in}
\abstract{
Recent measurements of $R_{D^*}$ have reduced tension with the Standard Model prediction. Taking all the present data into account, we obtain the values of the Wilson coefficients of each new physics four-fermion operator of a given Lorentz structure. 
We find that the combined data rule out most of the solutions based on scalar/pseudoscalar operators.  By studying the inter-relations between different solutions, we find that there are only four allowed solutions, which are based on operators with $(V-A)$, linear combination of $(V-A)$ and $(V+A)$, tensor and linear combination of scalar/pseudoscalar and tensor structure. We demonstrate that the need for new physics is driven by those measurement of $R_D$ and $R_{D^*}$ where the $\tau$ lepton is not studied.
 Further, we show that new physics only in $b\rightarrow c\,\mu\,\bar{\nu}$ is not compatible with the full set of observables in the decays $B\rightarrow Dl\bar{\nu}$ and $B\rightarrow D^*l\bar{\nu}$. 
}

\begin{document}
\maketitle
\flushbottom

\section{Introduction} 

 Over the last few years, there have been measurements of several 
observables in the $B$ meson sector which exhibit deviations from their 
standard model (SM) predictions.  These tensions with the SM are at the level of 2-4 $\sigma$ and can be considered as signatures of possible physics beyond the SM. Two such observables are 
\begin{equation}
R_{D} = \frac{\Gamma(B\rightarrow D\,\tau\,\bar{\nu})}{ \Gamma(B\rightarrow D\, e/\mu \, \bar{\nu})}\,,\quad
R_{D^{*}} = \frac{\Gamma(B\rightarrow D^{*}\,\tau\,\bar{\nu})}{ \Gamma(B\rightarrow D^{*}\, e/\mu \, \bar{\nu})}\,.
\end{equation}
 The decays $B\rightarrow D/D^{*}\, l \, \bar{\nu}$ are induced by the quark level transition $b\rightarrow c\, l \, \bar{\nu}$ which occurs at the tree level within the SM. The evidence for an excess in  $R_{D}$/$R_{D^{*}}$ is provided by a series of measurements by BaBar \cite{Lees:2012xj,Lees:2013uzd}, Belle \cite{Huschle:2015rga,Sato:2016svk,Hirose:2016wfn} and LHCb 
 \cite{Aaij:2015yra,Aaij:2017uff,Aaij:2017deq} collaborations.  The SM prediction for $R_D$ is $0.300\pm 0.008$~\cite{Aoki:2016frl} whereas  the world average of its measured values is $0.407\pm 0.039\pm 0.024$. For $R_{D^*}$ the corresponding numbers are $0.252\pm 0.003$~\cite{Fajfer:2012vx} and $0.304\pm 0.013\pm 0.007$ respectively. In the world average values, taken from~\cite{average}, the first error is statistical and the second error is systematic. Theoretical predictions of $R_{D}$/$R_{D^{*}}$ have been updated recently, using different approaches, see for e.g., refs. ~\cite{Bigi:2016mdz,Bernlochner:2017jka,Bigi:2017jbd,Jaiswal:2017rve,deBoer:2018ipi}.

The present average values of $R_{D}$ and $R_{D^*}$ exceed the SM predictions by $2.3\sigma$ and $3.4\sigma$ respectively. Including the $R_D/R_{D^*}$ correlations, the tension is at the level of $4.1\sigma$ \cite{average}. This discrepancy is an indication that the lepton flavor universality (LFU), predicted by the SM, is violated. Using the full data sample of $772 \times 10^6$ $\rm B\bar{B}$ pairs, the Belle collaboration has recently reported their measurement of $\tau$ polarization in the $B \to D^* \tau \bar{\nu}$ decay \cite{Hirose:2016wfn}. The measured value, $P_{\tau}(D^*) = - 0.38 \pm 0.51 ^{+0.21}_{-0.16}$, is consistent with its SM prediction of $-0.497\pm0.013$ \cite{Tanaka:2012nw}.

Very recently the LHCb collaboration has measured a new ratio related to the quark level transition $b\rightarrow c\, l\, \bar{\nu}$ \cite{lhcb-new}
\begin{equation}
R_{J/\psi} = \frac{\Gamma(B_c\rightarrow J/\psi \, \tau \, \bar{\nu})}{\Gamma(B_c\rightarrow J/\psi \, \mu\, \bar{\nu})} = 0.71\pm 0.17(stat.) \pm 0.18(syst.),
\end{equation}
which is $1.7\sigma$ higher than its SM prediction $0.289\pm 0.010$ \cite{Dutta:2017xmj}. Thus this measurement reinforces the idea of LFU violation in the $b\rightarrow c\, l \, \bar{\nu}$ sector.

The $R_D/R_{D^*}$ anomalies have been studied in various model dependent and model independent approaches, see for example refs. \cite{Fajfer:2012vx,Tanaka:2012nw,Fajfer:2012jt,Alonso:2015sja,Ivanov:2017mrj,Datta:2012qk,Biancofiore:2013ki,Duraisamy:2013kcw,Duraisamy:2014sna,
Sakaki:2014sea,Freytsis:2015qca,Becirevic:2016hea,Alonso:2016gym,Alok:2016qyh,Ivanov:2016qtw,Ligeti:2016npd,Bardhan:2016uhr,Kim:2016yth,Dutta:2016eml,Bhattacharya:2016zcw,Alonso:2016oyd,Alonso:2017ktd,Jung:2018lfu,Colangelo:2018cnj} for model independent analyses, refs. \cite{Sakaki:2013bfa,Fajfer:2015ycq,Bauer:2015knc,Barbieri:2015yvd,Dorsner:2016wpm,Li:2016vvp,Sahoo:2016pet,Bhattacharya:2016mcc,Barbieri:2016las,
Chen:2017hir,Crivellin:2017zlb,Alok:2017jaf,Calibbi:2017qbu} for leptoquark models, refs. \cite{Crivellin:2012ye,Celis:2012dk,Crivellin:2015hha,Wang:2016ggf,Celis:2016azn,Ko:2017lzd,Iguro:2017ysu,Biswas:2018jun,Martinez:2018ynq} for models with extra scalars, refs. \cite{Greljo:2015mma,Boucenna:2016wpr,Matsuzaki:2017bpp,Asadi:2018wea} for extended vector boson models, and refs. \cite{Das:2016vkr,Deshpand:2016cpw,Cvetic:2017gkt,Aloni:2017eny,Megias:2017ove,Altmannshofer:2017poe,Feruglio:2017rjo,Choudhury:2017qyt,Cline:2017ihf} for other new physics (NP) scenarios. In ref. \cite{Freytsis:2015qca}, all possible four-fermion operators for the $b\rightarrow c\,\tau\,\bar{\nu}$ decay were identified and the constraints on their Wilson coefficients (WCs) were derived by fitting them to the $R_{D}$ and $R_{D^*}$ data. The solutions with pseudoscalar operators are subject to an additional constraint from  
the leptonic decay $B_c \to \tau\,\bar{\nu}$~\cite{Li:2016vvp,Alonso:2016oyd}. This constraint rules out most of the NP solutions containing pseudoscalar operators.
The effect of $R_{J/\psi}$ anomaly on the NP in 
$b \to c\, \tau\, \bar{\nu}$ sector is considered in refs. 
\cite{Watanabe:2017mip,Chauhan:2017uil,Dutta:2017wpq,Tran:2018kuv,
Wei:2018vmk,Rui:2018kqr}.

In this paper we do a re-analysis of all available data on $b\rightarrow c\,\tau\,\bar{\nu}$ transition to find the allowed NP solutions and the values of corresponding WCs. We include the new data on $R_{D^*}$ from Belle~\cite{Hirose:2016wfn} and LHCb~\cite{Aaij:2017uff,Aaij:2017deq}, which has a smaller deviation
from the SM, and the recent data on $R_{J/\psi}$. We discuss the inter-relations between different solutions and show that there are essentially four NP solutions. We comment on ways of distinguishing between the four allowed solutions using angular asymmetries and polarization fractions. Further, we analyse the impact of individual measurements from BaBar, Belle and LHCb on the NP in $b \to c\, \tau\, \bar{\nu}$ sector. We also consider the scenario with NP in $b \to c\, \mu\, \bar{\nu}$ amplitude and show that such a scenario is ruled out by data. 
 
The paper is organized as follows. In Section II, we describe our calculation and present our results. 
In Section III, we study the impact of individual measurements from BaBar, Belle and LHCb experiments on the NP in $b \to c\, \tau\, \bar{\nu}$ transition. In section IV, we discuss the scenario of NP only in $b \to c\, \mu\, \bar{\nu}$.
In Section V, we present our conclusions.

\section{Calculation and results}


 The most general effective Hamiltonian for $b\rightarrow c\tau\bar{\nu}$ transition, containing all possible Lorentz structures, is \cite{Freytsis:2015qca}
\begin{equation}
H_{eff}= \frac{4 G_F}{\sqrt{2}} V_{cb}\left[O_{V_L} + \frac{\sqrt{2}}{4 G_F V_{cb}} \frac{1}{\Lambda^2} \left\lbrace \sum_i \left(C_i O_i +
 C^{'}_i O^{'}_i + C^{''}_i O^{''}_i \right) \right\rbrace \right],
\label{effH}
\end{equation}
where $G_F$ is the Fermi coupling constant and $V_{cb}$ is the Cabibbo-Kobayashi-Maskawa (CKM) 
matrix element. Here $O_{V_L}$ is the SM operator which has the 
usual $(V-A) * (V -A)$ structure. The explicit forms of the four-fermion operators $O_i$, $O^{'}_i$ and $O^{''}_i$ are given in the table \ref{tab1}. In writing the above Hamiltonian we assume that the neutrino is always left chiral. Hence we do not consider operators containing $\nu_R$. Solutions to $R_D$/$R_{D^*}$ problem involving right chiral neutrinos are discussed in refs.~\cite{Asadi:2018wea,Greljo:2018ogz,Robinson:2018gza}.
The constants $C_i$, $C^{'}_i$ and $C^{''}_i$ are the respective Wilson coefficients of the NP operators in which NP effects are encoded.  The table also gives the Fierz transformed forms of primed and double primed operators in terms of the unprimed operators. For later convenience we define $(2\sqrt{2} G_F V_{cb}\Lambda^2)^{-1} \equiv \alpha$. We set the new physics scale $\Lambda$ to be 1 TeV, which leads to $\alpha = 0.749$.

\begin{table}[h!]
\centering
\begin{tabular}{|l|ccc|}
  \hline\hline
	& Operator & & Fierz identity\\
  \hline
$O_{V_L}$   & $(\bar{c} \gamma_\mu P_L b)\,(\bar{\tau} \gamma^\mu P_L \nu)$ & & \\
$O_{V_R}$   & $(\bar{c} \gamma_\mu P_R b)\,(\bar{\tau} \gamma^\mu P_L \nu)$ & & \\
  $O_{S_R}$   & $(\bar{c} P_R b)\,(\bar{\tau} P_L \nu)$ & & \\
  $O_{S_L}$   & $(\bar{c} P_L b)\,(\bar{\tau} P_L \nu)$ & &\\
  $O_T$       & $(\bar{c}\sigma^{\mu\nu}P_L b)\,(\bar{\tau}\sigma_{\mu\nu}P_L \nu)$ & &\\[2pt]
  \hline
 $O^{'}_{V_L}$ &$(\bar{\tau} \gamma_\mu P_L b)\,(\bar{c} \gamma^\mu P_L \nu)$ &
   $\longleftrightarrow$ &$O_{V_L}$\\
$O^{'}_{V_R}$  & $(\bar{\tau} \gamma_\mu P_R b)\,(\bar{c} \gamma^\mu P_L \nu)$ &
    $\longleftrightarrow$ & $-2O_{S_R}$ \\
 $O^{'}_{S_R}$  & $(\bar{\tau} P_R b)\,(\bar{c} P_L \nu)$ &
    $\longleftrightarrow$ & $-\frac{1}{2}O_{V_R}$ \\
  $O^{'}_{S_L}$  & $(\bar{\tau} P_L b)\,(\bar{c} P_L \nu)$ &
    $\longleftrightarrow$ & $-\frac{1}{2}O_{S_L} - \frac{1}{8}O_T$\\
$O^{'}_T$      & $(\bar{\tau}\sigma^{\mu\nu}P_L b)\,(\bar{c}\sigma_{\mu\nu}P_L \nu)$ &
    $\longleftrightarrow$ & $-6O_{S_L} + \frac{1}{2}O_T$ \\[2pt]
  \hline
 $O^{''}_{V_L}$ & $(\bar{\tau} \gamma_\mu P_L c^c)\,(\bar{b}^c \gamma^\mu P_L \nu)$ &
    $\longleftrightarrow$ & $-O_{V_R}$  \\
  $O^{''}_{V_R}$ & $(\bar{\tau} \gamma_\mu P_R c^c)\,(\bar{b}^c \gamma^\mu P_L \nu)$ &
    $\longleftrightarrow$ & $-2O_{S_R}$ \\
$O^{''}_{S_R}$ &$(\bar{\tau} P_R c^c)\,(\bar{b}^c P_L \nu)$ &
$\longleftrightarrow$ & $\frac{1}{2}O_{V_L}$ \\
$O^{''}_{S_L}$ & $(\bar{\tau} P_L c^c)\,(\bar{b}^c P_L \nu)$ &
    $\longleftrightarrow$ & $-\frac{1}{2}O_{S_L} + \frac{1}{8}O_T$ \\
 $O^{''}_T$     & $(\bar{\tau}\sigma^{\mu\nu}P_L c^c)\,(\bar{b}^c\sigma_{\mu\nu}P_L \nu)$ &
    $\longleftrightarrow$ & $-6O_{S_L} - \frac{1}{2}O_T$  \\[2pt]
  \hline\hline
\end{tabular}
\caption{All possible four-fermion operators that can contribute to $ b \to
c \tau\bar{\nu}$ transition.}
\label{tab1}
\end{table}

 The primed and double primed operators are products of quark-lepton bilinears. They arise naturally in models containing leptoquarks~\cite{Davidson:1993qk}. Models containing leptoquarks of charge $2/3$, for example the model in ref.~\cite{Fajfer:2015ycq}, give rise to primed operators. The double primed operators occur due to the exchange of charge $1/3$ leptoquark, such as those in the model of ref.~\cite{Davidson:2010uu}. For this reason we have explicitly included these operators in our analysis eventhough they are linear combinations of unprimed operators. For a complete discussion of the properties of various leptoquark models, please see ref.~\cite{Dorsner:2016wpm}.

All possible NP WCs which provided good fit to the $R_{D}$ and $R_{D^*}$ data were calculated in ref. \cite{Freytsis:2015qca}. However the world averages for $R_{D^*}$  have  shifted owing to updates from Belle \cite{Hirose:2016wfn} and LHCb \cite{Aaij:2017uff}. Hence it is worth redoing the analysis with the new world average \cite{average} to see changes to the allowed solutions obtained in  \cite{Freytsis:2015qca}.  
Using the effective Hamiltonian given in Eq.~(\ref{effH}), we compute the observables $R_D$, $R_{D^*}$ and $P_{\tau}$ as functions of the various Wilson coefficients. By fitting these expressions to the measured values of the observables, we obtain the values of WCs which
are consistent with the data. Here we consider either one NP operator or a combination of two similar operators (for example [$O_{V_L}$, $O_{V_R}$], [$O_{S_L}$, $O_{S_R}$], [$O^{'}_{V_L}$, $O^{'}_{V_R}$] and [$O^{''}_{S_L}$, $O^{''}_{S_R}$]) at a time 
while making the fit to the experimental observables. 

First we fit the NP predictions to the three observables $R_D$, $R_{D^*}$ and $P_{\tau}$. The corresponding $\chi^2$ is defined as
\begin{eqnarray}
\chi^2(C^{\rm{eff}}_i)&=&\sum_{m,n= R_D, R_{D^*}}\left(O^{th}(C^{\rm{eff}}_i)-O^{exp}\right)_{m}\left(V^{exp}+V^{SM}\right)^{-1}_{mn}\left(O^{th}(C^{\rm{eff}}_i)-O^{exp}\right)_{n}\nonumber\\
& &+ \frac{(P_{\tau}^{th}(C^{\rm{eff}}_i)-P_{\tau}^{exp})^2}{\sigma^2_{P_{\tau}}}.
\label{chi2}
\end{eqnarray}
Here $O^{th}(C^{\rm{eff}}_i)$ are the theoretical predictions for $R_D$, $R_{D^*}$ which depend upon the effective NP WCs $C^{\rm{eff}}_i$. $O^{exp}$ are the corresponding experimental measurements. $V^{exp}$ and $V^{SM}$ are the experimental and SM covariance matrices in the $R_D$, $R_{D^*}$ space, respectively.  The matrix $V^{exp}$ includes the correlation in the combined experimental determination of $R_D$ and $R_{D^*}$.
In eq.~(\ref{chi2}), $\sigma_{P_{\tau}}$ is the uncertainty in the measurement of $P_{\tau}$. The expressions for various $C^{\rm{eff}}_i$, as linear combinations of $C_i$, $C^{'}_i$ and $C^{''}_i$, are defined below in eq.~(\ref{Cieff}).
\begin{eqnarray}
C^{\rm{eff}}_{V_L} &=&  \alpha \left(C_{V_L} + C^{'}_{V_L} + 0.5C^{''}_{S_R}\right), \nonumber \\
C^{\rm{eff}}_{V_R} &=& \alpha \left(C_{V_R} - 0.5C^{'}_{S_R} - C^{''}_{V_L}\right), \nonumber \\
C^{\rm{eff}}_{S_L} &=& \alpha \left(C_{S_L} -0.5C^{'}_{S_L}- 6C^{'}_{T}-0.5C^{''}_{S_L}-6C^{''}_{T}\right), \nonumber \\
C^{\rm{eff}}_{S_R} &=& \alpha \left(C_{S_R}-2C^{'}_{V_R}-2C^{''}_{V_R}\right), \nonumber \\
C^{\rm{eff}}_T &=& \alpha \left(C_T - 0.125C^{'}_{S_L}+0.5 C^{'}_T+0.125C^{''}_{S_L}-0.5C^{''}_T\right).
\label{Cieff}
\end{eqnarray}
The expressions for $R^{th}_D$, $R^{th}_{D^*}$ and $P^{th}_{\tau}$ in terms of $C^{\rm{eff}}_i$ are  
\begin{eqnarray}
R^{th}_D(C^{\rm{eff}}_i) &=& 0.297\mid 1+C^{\rm{eff}}_{V_L}+C^{\rm{eff}}_{V_R}\mid^2 + 0.398\mid C^{\rm{eff}}_{S_L}+C^{\rm{eff}}_{S_R}\mid^2 + 0.140\mid C^{\rm{eff}}_T\mid^2 \nonumber \\
& & + 0.509 Re\left[\left(1+C^{\rm{eff}}_{V_L}+C^{\rm{eff}}_{V_R}\right)\left(C^{*\rm{eff}}_{S_L}+C^{*\rm{eff}}_{S_R}\right)\right] \nonumber \\
& &+ 0.244 Re\left[\left(1+C^{\rm{eff}}_{V_L}+C^{\rm{eff}}_{V_R}\right)C^{*\rm{eff}}_{T}\right],
\end{eqnarray}
\begin{eqnarray}
R^{th}_{D^*}(C^{\rm{eff}}_i) & =& 0.253\left(\mid 1+C^{\rm{eff}}_{V_L}\mid^2 + \mid C^{\rm{eff}}_{V_R}\mid^2\right) - 0.449 Re\left[\left(1+C^{\rm{eff}}_{V_L}\right)C^{*\rm{eff}}_{V_R}\right]\nonumber\\
& &+ 0.011\mid C^{\rm{eff}}_{S_R}-C^{\rm{eff}}_{S_L}\mid^2 + 3.077\mid C^{\rm{eff}}_{T}\mid^2 \nonumber \\
& & + 0.030 Re\left[\left(1+C^{\rm{eff}}_{V_L}-C^{\rm{eff}}_{V_R}\right)\left(C^{*\rm{eff}}_{S_R}-C^{*\rm{eff}}_{S_L}\right)\right]\nonumber\\
& & - 1.055 Re\left[\left(1+C^{\rm{eff}}_{V_L}\right)C^{*\rm{eff}}_{T}\right] + 1.450 Re\left[C^{\rm{eff}}_{V_R}C^{*\rm{eff}}_T\right],
\end{eqnarray}
\begin{eqnarray}
P^{th}_{\tau}(C^{\rm{eff}}_i) &=& \left\lbrace 0.252\mid C^{\rm{eff}}_{S_L} - C^{\rm{eff}}_{S_R}\mid^2 + 4.089\mid C^{\rm{eff}}_T\mid^2 - 
   2.985 \left(\mid 1 + C^{\rm{eff}}_{V_L}\mid^2 + \mid C^{\rm{eff}}_{V_R}\mid^2\right)\right.\nonumber\\
& &  +8.298 Re\left[C^{*\rm{eff}}_T \left(1 + C^{\rm{eff}}_{V_L}\right)\right] + 
   0.716 Re\left[\left(C^{*\rm{eff}}_{S_R}-C^{*\rm{eff}}_{S_L}\right) \left(1 + C^{\rm{eff}}_{V_L} - C^{\rm{eff}}_{V_R}\right)\right] \nonumber\\
& & \left. -11.410 Re\left[C^{*\rm{eff}}_T C^{\rm{eff}}_{V_R}\right] + 5.136 Re\left[\left(1 + C^{\rm{eff}}_{V_L}\right) C^{*\rm{eff}}_{V_R}\right]\right\rbrace /
   \left\lbrace 0.252 \mid C^{\rm{eff}}_{S_L} - C^{\rm{eff}}_{S_R}\mid^2 \right.\nonumber\\
   && +  72.609 \mid C^{\rm{eff}}_T\mid^2 + 5.983 \left(\mid 1 + C^{\rm{eff}}_{V_L}\mid^2 + \mid C^{\rm{eff}}_{V_R}\mid^2 \right) - 
   24.894 Re\left[C^{*\rm{eff}}_T \left(1 + C^{\rm{eff}}_{V_L}\right)\right]\nonumber\\  
   &&  + 0.716 Re\left[\left(C^{*\rm{eff}}_{S_R} - C^{*\rm{eff}}_{S_L}\right) \left(1 + C^{\rm{eff}}_{V_L} - C^{\rm{eff}}_{V_R}\right)\right] + 
   34.232 Re\left[C^{*\rm{eff}}_T C^{\rm{eff}}_{V_R}\right] \nonumber\\
   & & \left. - 10.625 Re\left[\left(1 + C^{\rm{eff}}_{V_L}\right) C^{*\rm{eff}}_{V_R}\right]\right\rbrace.
\end{eqnarray}

The  $B\rightarrow D/D^{*}\, l\,  \bar{\nu}$ decay distributions depend upon hadronic form-factors. The determination of these form-factors relies heavily on HQET techniques.  In this work we use the HQET form factors in the form parametrized by Caprini {\it et al.} \cite{Caprini:1997mu}. The parameters for $B\rightarrow D$ decay are determined from the lattice  QCD \cite{Aoki:2016frl} calculations and we use them in our analyses. For $B\rightarrow D^*$ decay, the HQET parameters are extracted using data from Belle and BaBar experiments along with the inputs from lattice. In this work, the numerical values of these parameters are taken from refs. \cite{Bailey:2014tva} and  \cite{Amhis:2016xyh}.
 
 Here our $\chi^2$ includes experimental and SM theory uncertainties. For a given NP operator, the most likely value of its WC is obtained by minimizing the $\chi^2$.  For this minimization, we use the $\tt MINUIT$ library \cite{minuit1,minuit2}. We find that the values of $\chi^2_{min}$ fall into two disjoint ranges ($\lesssim 1$ and $\gtrsim 5$). The latter range occurs for NP structures such as $C_{V_R}$, $C_{S_R}$, $C'_{V_R}$ etc. The WCs of NP solutions with $\chi^2_{min}\lesssim 1$ are listed in table \ref{tab2}. 
\begin{table}[htbp]
\centering
\tabcolsep 6pt

\begin{tabular}{|c|c|c|c|}
\hline\hline
Coefficient(s)  &  Best fit value(s)  & $\chi^2_{min}$ & \emph{pull} \\
\hline
$C_{V_L}$  &  $0.146 $ & 0.92 & 4.32 \\
$C_{V_L}$  &  $-2.818 $ & 0.92 & 4.32\\
\hline
$C_{S_L}$  & $-1.917$ & 1.03& 4.31\\
\hline
$C_T$  &  $0.514 $   & 0.66& 4.35\\
\hline
$C'_{V_L}$ & $0.146$& 0.92 & 4.32\\
$C'_{V_L}$ & $-2.818$& 0.92 & 4.32\\
\hline
$C''_{S_L}$ & $3.544$ & 1.07 & 4.30\\
$C''_{S_L}$ & $-0.518$ & 0.07 & 4.42\\
$C''_{S_R}$ & $-5.636$ & 0.92 & 4.32\\
$C''_{S_R}$ & $0.292$ & 0.92 & 4.32\\
\hline
$(C_{V_L},\, C_{V_R})$ & $(-1.282,1.511)$ & 0.04& 4.42 \\
$(C_{V_L},\, C_{V_R})$ & $(0.175,0.053)$ & 0.04 & 4.42\\
$(C_{V_L},\, C_{V_R})$ & $(-1.389,-1.511)$ & 0.04 & 4.42\\
$(C_{V_L},\, C_{V_R})$ & $(-2.846,-0.053)$ & 0.04 & 4.42\\
\hline 
$(C_{S_L},\, C_{S_R})$  &  $(-0.659,0.912)$ & 0.05 & 4.42\\
$(C_{S_L},\, C_{S_R})$  &  $(2.813,-2.560)$  & 0.05 & 4.42\\
$(C_{S_L},\, C_{S_R})$ & $(-1.766, -0.195)$ & 0.05 & 4.42\\
\hline
$(C'_{V_L},\, C'_{V_R})$  &  $(0.120, -0.062)$ & 0.03& 4.42 \\
$(C'_{V_L},\, C'_{V_R})$  &  $(0.194, 0.997)$  & 0.18& 4.40 \\
$(C'_{V_L},\, C'_{V_R})$ & $(-2.792, 0.062)$ & 0.03 & 4.42\\
$(C'_{V_L},\, C'_{V_R})$ & $(-2.866, -0.997)$ & 0.18 &4.40\\
\hline
$(C'_{S_L},\, C'_{S_R})$ & $(-1.561, 1.221)$ & 0.15& 4.41 \\
$(C'_{S_L},\, C'_{S_R})$ & $(3.747, -2.746)$ & 0.86 & 4.33\\
\hline
$(C''_{V_L},\, C''_{V_R})$ & $(0.116, -0.185)$ & 0.01& 4.43\\ 
$(C''_{V_L},\, C''_{V_R})$ & $(0.216, 0.949)$ & 0.17& 4.40\\
\hline
$(C''_{S_L},\, C''_{S_R})$  & $(-0.655, -0.090)$  & 0.02 & 4.42\\
$(C''_{S_L},\, C''_{S_R})$  &  $(3.453, 0.180)$  & 0.87& 4.33 \\
$(C''_{S_L},\, C''_{S_R})$  &  $(0.655, -5.253)$  & 0.02& 4.42\\
$(C''_{S_L},\, C''_{S_R})$  &  $(-3.453, -5.523)$ & 0.87 & 4.33\\
\hline\hline
\end{tabular}
\caption{Best fit values of the WCs of NP operators at $\Lambda = 1$ TeV for the measurements of  $R_D$, $R_{D^*}$ and $P_{\tau}$. Here we list solutions with $\chi^2_{min}\lesssim 1$. For the SM, we have $\chi^2_{SM} = 19.61$. The pull values are calculated using \emph{pull} = $\sqrt{\chi^2_{SM}-\chi^2_{min}}$.}
\label{tab2}
\end{table}
\begin{table}[htbp]
\centering
\tabcolsep 6pt

\begin{tabular}{|c|c|l|}
\hline\hline
Coefficient(s)  & Best fit value(s)& $Br(B_c\rightarrow \tau\bar{\nu})$  \\
\hline
$C_{V_L}$  &  $0.149 $ & $2.66\times 10^{-2}$\\
\hline
$C_{S_L}$  & $-1.920$  & 3.04  \\

\hline
$C_T$  &  $0.516 $   & $2.15\times 10^{-2}$  \\

\hline
$C''_{S_L}$ & $3.551$ & 2.64  \\
$C''_{S_L}$ & $-0.526$ & $5.19\times 10^{-3}$   \\
\hline
$(C_{V_L},C_{V_R})$& $(-1.286, 1.512)$& $2.58\times 10^{-2}$\\

\hline 
$(C_{S_L},\, C_{S_R})$  &  $(-0.682,0.933)$ & 2.22\\
$(C_{S_L},\, C_{S_R})$  &  $(2.833,-2.583)$  & 19.01\\
\hline
$(C'_{V_L},\, C'_{V_R})$  &  $(0.124, -0.058)$ &  $6.60\times 10^{-2}$\\
$(C'_{V_L},\, C'_{V_R})$  &  $(0.198, 0.997)$  & 2.22 \\
\hline
$(C'_{S_L},\, C'_{S_R})$ & $(-1.561, 1.231)$ &  $1.89\times 10^{-1}$  \\
$(C'_{S_L},\, C'_{S_R})$ & $(3.750, -2.739)$ & 2.42\\
\hline
$(C''_{V_L},\, C''_{V_R})$ & $(0.120, -0.186)$ & $2.21\times 10^{-1}$  \\
$(C''_{V_L},\, C''_{V_R})$ & $(0.221, 0.948)$ &  1.97\\
\hline
$(C''_{S_L},\, C''_{S_R})$  & $(-0.643, -0.076)$  & $1.55\times 10^{-2}$ \\
$(C''_{S_L},\, C''_{S_R})$  &  $(3.436, 0.219)$  & 2.52 \\
\hline\hline
\end{tabular}
\caption{ Best fit values of the coefficients of new physics operators  at $\Lambda = 1$ TeV for the measurements of $R_D$, $R_{D^*}$, $R_{J/\psi}$ and $P_{\tau}$. Here we allow only those solutions for which $\chi^2_{min}\leq 4.8$. We also provide the predictions of branching ratio of $B_c\rightarrow \tau\bar{\nu}$ decay for each solution.}
\label{tab3}
\end{table}

Comparing the results of table~\ref{tab2} with those of \cite{Freytsis:2015qca}, it can be seen that more new physics solutions are now allowed. 
 In table~\ref{tab2}, the solutions with $C'_{V_L}$ and $C''_{S_R}$ are degenerate with the $C_{V_L}$ solutions, because $O_{V_L}$ is the Fierz transform of both $O'_{V_L}$ and $2\,O''_{S_R}$. For these three NP operators, the values of $R_D$ and $R_{D^*}$ are proportional to $[1+\alpha C_{V_L}]^2$. Hence, we get two degenerate solutions with equal and opposite values for $(1+\alpha C_{V_L})$. If both $C_{V_L}$ and $C_{V_R}$ are non-zero, the theoretical expression for $R_D$ is proportional to $[1+\alpha (C_{V_L}+C_{V_R})]^2$ and that for $R_{D^*}$ depends on $[1+\alpha(C_{V_L}+C_{V_R})]^2$ and $[1+\alpha (C_{V_L}-C_{V_R})]^2$. There are four combinations of ($C_{V_L}, C_{V_R}$) which have the same values for the above two functions and hence are degenerate. For two of these combinations the value of $C_{V_R}$ is negligibly small and the value of $C_{V_L}$ is close to that of the $O_{V_L}$ solution. The other two combinations have equal and opposite values of $[1+\alpha(C_{V_L}+C_{V_R})]$ and $[1+\alpha (C_{V_L}-C_{V_R})]$. Hence, there is essentially only one NP solution with non-zero $C_{V_L}$ and $C_{V_R}$. Similar explanations can be found for other degeneracies. Among multiple degenerate solutions, we keep only one solution in further analysis. 

We now include $R_{J/\psi}$ in our fit. The expression for $\chi^2$ will have an additional term 
\begin{equation}
\frac{\left(R_{J/\psi}^{th}(C^{\rm{eff}}_i)-R_{J/\psi}^{exp}\right)^2}{\sigma^2_{R_{J/\psi}}},
\end{equation}
where $\sigma_{R_{J/\psi}}$ is the experimental uncertainty in $R_{J/\psi}$. 
 The expression of $R^{th}_{J/\psi}$
\begin{eqnarray}
R^{th}_{J/\psi}(C^{\rm{eff}}_i) & =& 0.289\left(\mid 1+C^{\rm{eff}}_{V_L}\mid^2 + \mid C^{\rm{eff}}_{V_R}\mid^2\right) - 0.559 Re\left[\left(1+C^{\rm{eff}}_{V_L}\right)C^{*\rm{eff}}_{V_R}\right]\nonumber\\
& &+ 0.014\mid C^{\rm{eff}}_{S_R}-C^{\rm{eff}}_{S_L}\mid^2 + 3.095\mid C^{\rm{eff}}_{T}\mid^2 \nonumber \\
& & + 0.041 Re\left[\left(1+C^{\rm{eff}}_{V_L}-C^{\rm{eff}}_{V_R}\right)\left(C^{*\rm{eff}}_{S_R}-C^{*\rm{eff}}_{S_L}\right)\right]\nonumber\\
& & - 1.421 Re\left[\left(1+C^{\rm{eff}}_{V_L}\right)C^{*\rm{eff}}_{T}\right] + 1.562 Re\left[C^{\rm{eff}}_{V_R}C^{*\rm{eff}}_T\right]
\end{eqnarray}
 The form factors for $B_c\rightarrow J/\psi$ transition and their uncertainties from ref.~\cite{Wen-Fei:2013uea} are used in the calculation of $R_{J/\psi}^{th}$. These form factors are calculated in perturbative QCD framework. After the inclusion of $R_{J/\psi}$ data,  the values of $\chi^2_{min}$ again fall into two disjoint ranges ($\leq 4.8$ and $\geq 7.5$). The WCs of NP solutions with $\chi^2_{min}\leq 4.8$ are listed in table \ref{tab3}. The comparison of the three observables fit with the four observables fit shows some remarkable features. Each solution with $\chi^2_{min}\lesssim 1$ in the three observables fit has a corresponding solution with $\chi^2_{min}\leq 4.8$ in the four observables fit. The WCs in the two cases are very close to each other. In addition, $\chi^2_{min}\gtrsim 5$ solutions of three observables fit all have $\chi^2_{min}\geq 7.5$ in four observables fit. Hence we conclude that the NP which can explain $R_D/R_{D^*}$ can also account for $R_{J/\psi}$. At present, there is no tension between $R_{J/\psi}$ measurement and the $R_D$ and $R_{D^*}$ measurements.

We now consider the constraint from the purely leptonic decay $B_c\rightarrow \tau\,\bar{\nu}$.
This decay is not subject to helicity suppression if the $b\rightarrow c\,\tau\,\bar{\nu}$ transition is through pseudo-scalar operators. The prediction for the partial width of the above decay, in such cases, is likely to be larger than the measured total decay width of $B_c$ meson. Hence the ratio 
\begin{equation}
Br(B_c\rightarrow \tau\bar{\nu}) = \frac{\Gamma(B_c\rightarrow \tau\bar{\nu})_{\rm NP}}{\Gamma(B_c\rightarrow {\rm all})_{\rm exp}}
\label{branch}
\end{equation} 
puts strong constraints on the allowed NP WCs. 
 The most general expression for the branching fraction of $B_c\rightarrow \tau\bar{\nu}$ is 
\begin{eqnarray}
Br(B_c\rightarrow \tau\bar{\nu}) &=& \frac{\vert V_{cb}\vert^2G^2_Ff^2_{B_c}m_{B_c}m^2_{\tau}\tau^{exp}_{B_c}}{8\pi}\left(1-\frac{m^2_{\tau}}{m^2_{B_c}}\right)^2\times \nonumber\\
& &  \left| 1+C^{\rm{eff}}_{V_L}-C^{\rm{eff}}_{V_R}+\frac{m^2_{B_c}}{m_{\tau}(m_b+m_c)}(C^{\rm{eff}}_{S_R}-C^{\rm{eff}}_{S_L})\right|^2
\end{eqnarray}
where $f_{B_c}= 434\pm 15$ MeV \cite{Colquhoun:2015oha} and  $\tau^{exp}_{B_c}=0.507\pm 0.009$ ps \cite{pdg}. The NP Wilson coefficients, $C^{\rm{eff}}_{V_L}$, $C^{\rm{eff}}_{V_R}$, $C^{\rm{eff}}_{S_L}$ and $C^{\rm{eff}}_{S_R}$, are defined in Eq.~(\ref{Cieff}). Here  $m_b$ and $m_c$ are the quark masses at the $\mu_b=m_b$ scale.

Table \ref{tab3} also lists predictions for $Br(B_c \to \tau\, \bar{\nu})$. In SM, the prediction for this branching ratio is reasonably small 
$( 2.15\times 10^{-2})$. We see from table~\ref{tab3} that some NP solutions, especially those with large pseudo-scalar couplings, predict this quantity to be greater than 1. Such solutions, obviously, are to be discarded. Recently, it is shown in ref.~\cite{Akeroyd:2017mhr} that LEP data imposes a constraint $Br(B_c \to \tau\, \bar{\nu})< 0.1$. This constraint rules out two of the new solutions $(C'_{S_L},\, C'_{S_R})= (-1.561, 1.231)$ and $(C''_{V_L},\, C''_{V_R})=(0.120, -0.186)$.

\begin{figure*}[h] 
\centering
\begin{tabular}{cc}
\includegraphics[width=60mm]{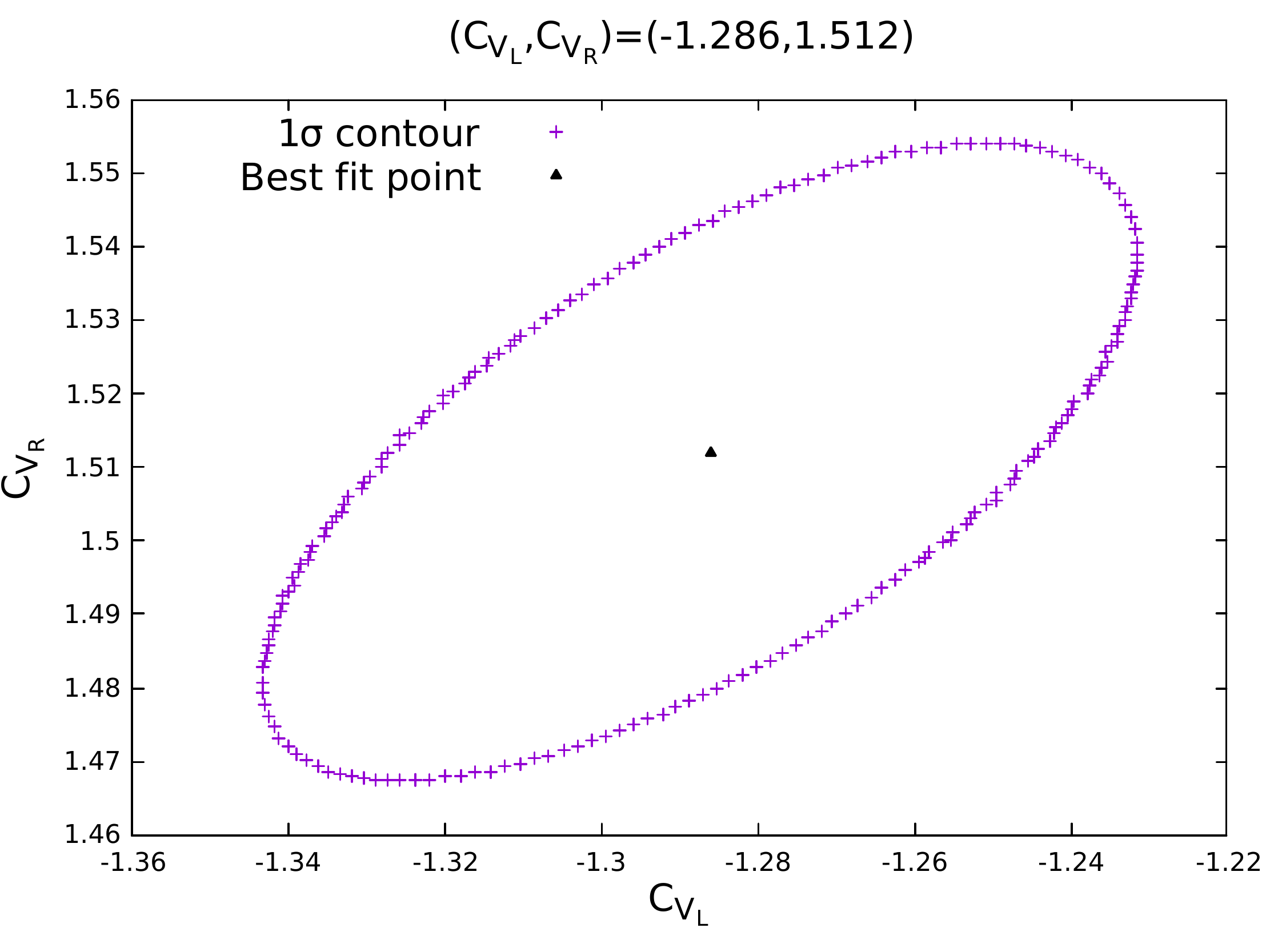}&
\includegraphics[width=60mm]{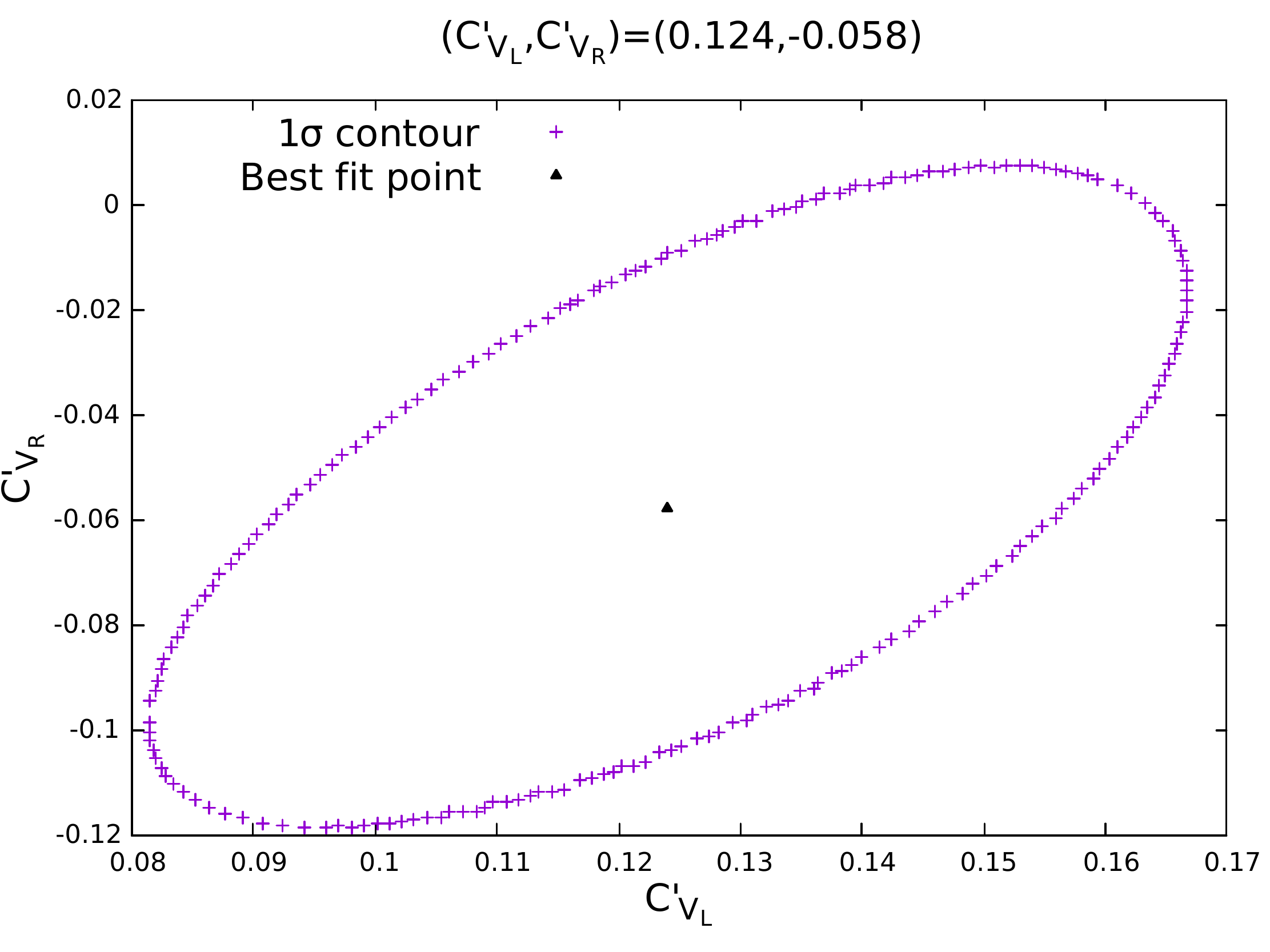}\\
\end{tabular}

\includegraphics[width=65mm]{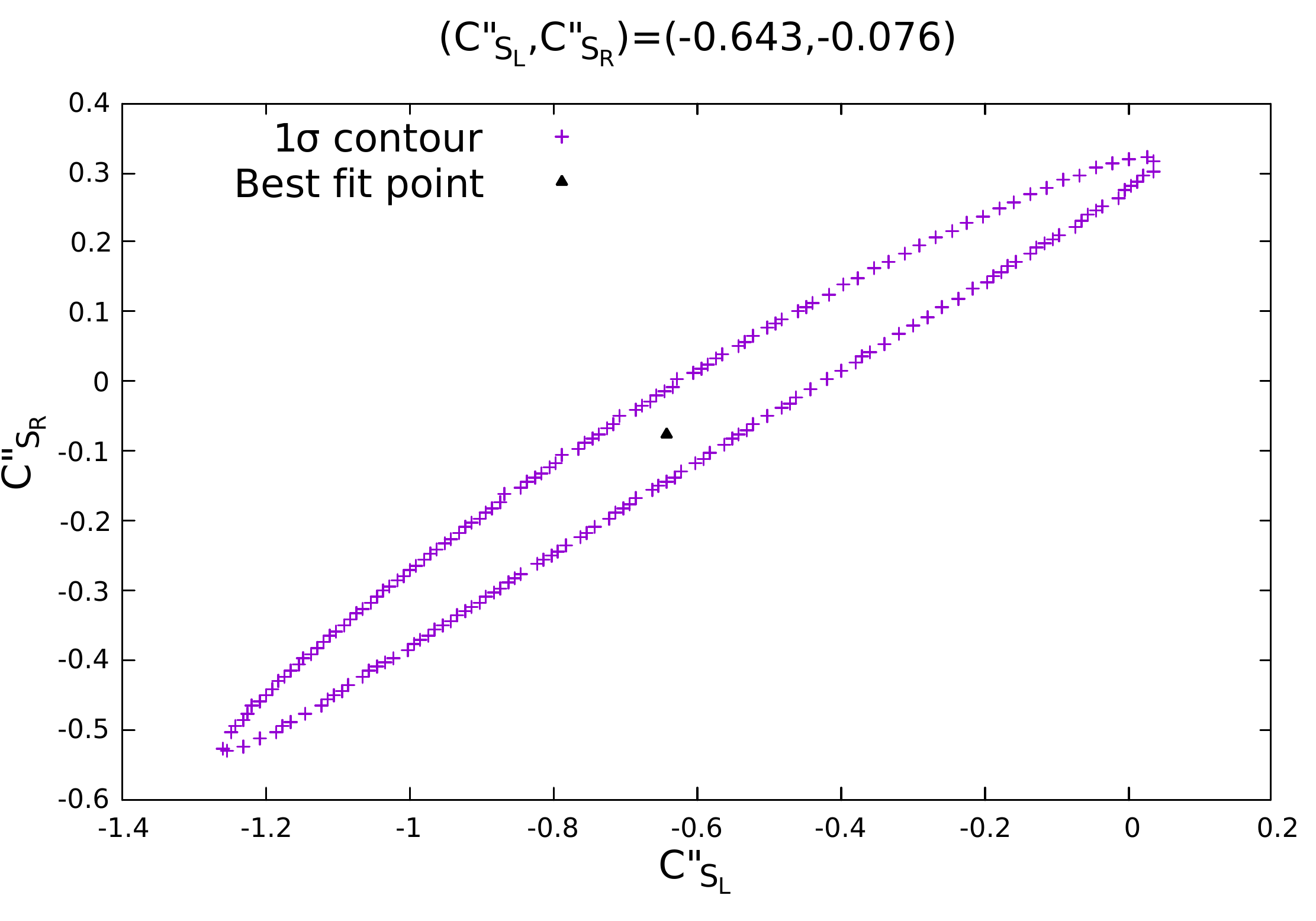}\\
\caption{The allowed $1\sigma$ ellipses for the two parameter solutions listed in table~\ref{tab4}.}
\label{ellipses}
\end{figure*}

The list of WCs of NP solutions which satisfy all the present experimental constraints is given in table~\ref{tab4}. Using the best fit values of the allowed solutions, we provide the predicted central values of the quantities used in the fit, i.e., $R_D, R_{D^*}, R_{J/\psi}$ and $P_{\tau}$, for each solution. This will allow us to see how close are the predictions of NP solutions to the experimental measurements. We also give the uncertanties on the  obtained values of WCs. The range for the uncertainty is calculated using the definition $\chi^2 (C_i) \leq \chi^2_{min}+1$. When two NP operators are considered together, the ranges of the corresponding WCs are correlated. These correlation ellipses are shown in fig.~\ref{ellipses} for the three allowed solutions with two NP operators.

Looking at the predictions in table~\ref{tab4} we make the following observations:
\begin{itemize}
\item There are only four NP solutions effectively because the fifth NP solution is essentially the same as the first NP solution and the sixth NP solution is essentially the same as the third NP solution. The value of $C'_{V_R}$ in the fifth solution is quite small and the value of $C'_{V_L}$ is close to the value of $C_{V_L}$ in the first solution. Since the Lorentz structure of $O'_{V_L}$ is the same as that of $O_{V_L}$, we can argue that these two solutions are essentially the same. Similarly, in the case of the sixth solution the value of $C''_{S_R}$ is very small and the value of $C''_{S_L}$ is close to that of the third solution.
\item Except for the tensor NP, all the other the predicted values of $R_{J/\psi}$ are about half of the central value of the experimental measurement, but are within $1.6\sigma$. Only for tensor NP the predicted $R_{J/\psi}$ is significantly smaller than the SM prediction.
\item The prediction for $P_{\tau}$ is markedly different only in the case of NP tensor couplings. In all other cases $P_{\tau}$ is predicted to be very close to the SM prediction. The reason for this is different for different cases. 
\begin{enumerate}
\item The $C_{V_L}$ solution has the same Lorentz structure as the SM and hence has the same prediction for $P_{\tau}$. 
\item In the case of $C''_{S_L}$ solution, the effective couplings of $O_{S_L}$ and $O_{T}$ operators are quite small and hence the prediction for $P_{\tau}$ remains close to the SM prediction. 
\item $P_{\tau}$ prediction for $O_{V_R}$ operator is the same as that of $O_{V_L}$ (SM) operator. Hence, a linear combination of these two operators also predicts $P_{\tau}$ to the same as the SM value. 
\end{enumerate}
\end{itemize}
\begin{table}[h!]
\centering
\tabcolsep 2pt
\begin{tabular}{|c|c|c|c|c|c|}
\hline\hline
NP type & Best fit value(s)  & $R_D$ & $R_{D^*}$ & $R_{J/\psi}$ & $P_{\tau}$ \\
\hline
SM  & $C_{i}=0$  &$0.297\pm 0.008$ &$0.253\pm 0.002$ & $0.289\pm 0.007$ &$-0.498\pm 0.004$ \\
\hline
$C_{V_L}$  &  $0.149 \pm 0.032$  &$0.366\pm 0.013$ &$0.313\pm 0.008$ & $0.358\pm 0.012$&$-0.498\pm 0.004$ \\
\hline
$C_T$  &  $0.516 \pm 0.015$   &$0.411\pm 0.012$&$0.304\pm 0.011$&$0.202\pm 0.016$&$+0.115\pm 0.013$ \\
\hline
$C''_{S_L}$ & $-0.526\pm 0.102$  &$0.400\pm 0.020$&$0.307\pm 0.012$&$0.359\pm 0.015$&$-0.484\pm 0.003$ \\

\hline
$(C_{V_L},C_{V_R})$& $(-1.286, 1.512)$& $0.405\pm 0.012$& $0.305\pm 0.003$& $0.348\pm 0.008$& $-0.499\pm 0.004$ \\

\hline
$(C'_{V_L},\, C'_{V_R})$  &  $(0.124, -0.058)$ & $0.406\pm 0.012$&$0.305\pm 0.003$ &$0.349\pm 0.009$ &$-0.484\pm 0.005$ \\
\hline
$(C''_{S_L},\, C''_{S_R})$  & $(-0.643, -0.076)$    &$0.408\pm 0.013$ & $0.305\pm 0.003$&$0.359\pm 0.005$ &$-0.477\pm 0.003$ \\

\hline\hline
\end{tabular}
\caption{ Best fit values of the coefficients of new physics operators at $\Lambda = 1$ TeV by making use of data of $R_D$, $R_{D^*}$, $R_{J/\psi}$ and $P_{\tau}$. Here we allow only those solutions for which $\chi^2_{min}\leq 4.8$ as well as $Br(B_c\rightarrow \tau\bar{\nu})< 0.1$. We also provide the predictions of $R_D$, $R_{D^*}$, $R_{J/\psi}$ and $P_{\tau}$ for each allowed solutions. We provide the $1\sigma$ error for 1-parameter fit and only the central values for 2-parameter fit. The $1\sigma$ ellipses for the three 2-parameter NP solutions are given in fig.~\ref{ellipses}. In the first row, the SM values of the observables come from our calculations by setting all NP couplings to zero.}
\label{tab4}
\end{table}

\section{Impact of individual measurements from BaBar, Belle and LHCb}

The analyses in the  previous section were done using the world averages of $R_D$ and $R_{D^*}$. The initial measurements of BaBar \cite{Lees:2012xj,Lees:2013uzd} and Belle \cite{Huschle:2015rga,Sato:2016svk}, where the $\tau$ lepton was not studied,  are well above the SM predictions. More recent measurements, where the experiments tried to reconstruct the $\tau$ lepton~\cite{Hirose:2016wfn,Aaij:2017uff,Aaij:2017deq}, are closer to the SM predictions. However, it must be emphasised that in each case the measured value is larger than the SM prediction. 

It is worthwhile to treat each individial measurement as a seperate data point to see how close the predictions of each NP solution are to the corresponding measurement. In  this section we do such an analysis by taking all the measurements related to $b\rightarrow c\tau\bar{\nu}$ as individual data points. There are ten such measurements including $P_{\tau}$ and $R_{J/\psi}$. But the measurements of $R_D$ and $R_{D^*}$ in BaBar~\cite{Lees:2013uzd} are correlated. The same is true for the Belle measurement~\cite{Huschle:2015rga}. In addition, the measurement of $R_{D^*}$ and $P_{\tau}$ by Belle~\cite{Hirose:2016wfn} is based on the same data set and hence are correlated.
We take the best fit value of NP WCs in each case and compute the $\chi^2$ for each individual measurement. For comparison we do this for SM also. The results are presented in table~\ref{tab5}. Since some of the measurements are correlated, the $\chi^2$ for these is computed, taking these correlations into account. Therefore we have only seven individual $\chi^2$ values listed in table~\ref{tab5}.
 
 From this table we can see the impact of various different measurements on the individual values of $\chi^2$ as well as the total $\chi^2$. First we note that the total $\chi^2$ is quite large for SM but $\leq 8$ for all NP solutions. In case of SM, except for BaBar~\cite{Lees:2013uzd}, all other individual $\chi^2$ values are $\lesssim 4$. For all NP solutions, except for tensor solution, $\chi^2(R_{J/\psi})\approxeq 2$ because the NP predictions are typically $1.5\sigma$ away from the measured central value. Since the tensor NP predicts $R_{J/\psi}$ significantly smaller than the SM prediction, the corresponding $\chi^2(R_{J/\psi})\approxeq 4$. All other individual $\chi^2$ values for all NP solutions satisfy $\chi^2\gtrsim 1$, meaning that the NP solutions do indeed provide a very good fit to each individual measurement. 
 
 For the three measurements, where the $\tau$ lepton is not reconstructed, the NP solutions lead to considerable reduction in the value of $\chi^2$ from the SM value. For BaBar~\cite{Lees:2013uzd}, this reduction is from $14$ to $\gtrsim 1$. In the case of Belle~\cite{Huschle:2015rga} and LHCb~\cite{Aaij:2015yra} the corresponding reduction is from $4$ to $\lesssim 1$. For the two data points~\cite{Hirose:2016wfn,Aaij:2017uff} where $\tau$ lepton is reconstructed, the difference between the $\chi^2$ values of the SM and the NP solutions is quite small. So we note that the need for NP is driven by those measurements where the $\tau$ lepton was not reconstructed.

\begin{table}[h]
\centering
\tabcolsep 2pt
\begin{tabular}{|c|c|c|c|c|c|c|c|c|c|}
\hline\hline
NP type & Best fit value(s) & $\chi^2$(All) &$\chi^2$(I) & $\chi^2$(II) & $\chi^2$(III)&  $\chi^2$(IV)& $\chi^2$(V)&$\chi^2$(VI) & $\chi^2$($R_{J/\psi}$)\\
\hline
 SM &$C_i = 0$& 28.88 & 14.40 & 4.27 & 2.27 & 0.12 & 4.22 & 0.77& 2.82 \\
\hline
$C_{V_L}$ & $0.152$&6.03& 1.76&0.29&0.14& 1.0 & 0.29 & 0.56& 1.96\\
\hline
$C_T$&  0.519&7.75&1.02 & 0.79& 0& 0.99& 0.49& 0.34& 4.08\\
\hline
$C''_{S_L}$&$-0.535$& 5.32& 1.18& 0.57& 0& 0.75& 0.48& 0.35& 1.95\\
\hline
$(C_{V_L},C_{V_R})$ & ($-1.293$, $1.511$)&5.46& 1.21& 0.54& 0& 0.77& 0.48& 0.34& 2.06\\ 
\hline
$(C'_{V_L},C'_{V_R})$& ($0.131$, $-0.052$)&5.41& 1.17& 0.58& 0& 0.75& 0.48& 0.34& 2.04\\
\hline
$(C''_{S_L},C''_{S_R})$&($-0.584$, $-0.032$)& 5.31& 1.18& 0.61& 0& 0.72& 0.51& 0.32& 1.95\\
\hline\hline
\end{tabular}
\caption{ List of $\chi^2$ values of each individual experiment for SM and for each NP solution. The notations for these experiments are: BaBar$\to$I, Belle'15$\to$II, Belle'16-I$\to$III, Belle'16-II$\to$IV, LHCb'15$\to$V, LHCb'17$\to$VI.}
\label{tab5}
\end{table}

\section{New physics in only $b\rightarrow c\,\mu\,\bar{\nu}$}
So far, we have discussed scenarios where NP contributes  only to $b\rightarrow c\,\tau\,\bar{\nu}$ transition.  It is also interesting to consider scenarios where the NP is {\bf not} in $b\rightarrow c\,\tau\,\bar{\nu}$ but  only in $b\rightarrow c\,\mu\,\bar{\nu}$.
 Such an assumption may give a good fit to $R_D$ and $R_{D^*}$ but is likely to disagree with other semi-leptonic decays of B mesons.
Belle has measured the two ratios in $b\rightarrow c\,l\,\nu$ sector \cite{Glattauer:2015teq}
\begin{equation}
R^{\mu/e}_D = \frac{\Gamma(B\rightarrow D\,\mu\,\nu)}{\Gamma(B\rightarrow D\,e\,\nu)} = 0.995\pm 0.022(stat.)\pm 0.039(syst.),
\label{rdme}  
\end{equation} 
and \cite{Abdesselam:2017kjf}
\begin{equation}
R^{e/\mu}_{D^*} = \frac{\Gamma(B\rightarrow D^* \,e \, \nu)}{\Gamma(B\rightarrow D^* \,\mu \, \nu)} = 1.04\pm 0.05 (stat.)\pm 0.01(syst.).
\label{rdstemu}
\end{equation}
These ratios are in agreement with their SM expectations. Any NP  only in $b\rightarrow c\,\mu\,\bar{\nu}$ will spoil this agreement. The small uncertainties in the above measurements will lead to negligibly small NP Wilson coefficients in the $b\rightarrow c \,\mu\,\bar{\nu}$ effective Lagrangian. If we take NP only in the muon sector and do a fit using the available experimental data of $R_D$, $R_{D^*}$, $R_{J/\psi}$, $R^{\mu/e}_D$ and $R^{e/\mu}_{D^*}$, we get $\chi^2_{min}$ only a little lower than the $\chi^2_{SM} = 23$. In particular, we get the $\chi^2_{min}> 19$ for one parameter fit and $\chi^2_{min}> 17.5$ for two parameter fit. Hence, we can conclude that NP in only  $b\rightarrow c\,\mu\,\bar{\nu}$ is not a viable explanation for $R_{D}$/$R_{D^*}$ anomaly. However,  it is possible to satisfy the constraints in eqs.~(\ref{rdme}) and~(\ref{rdstemu}) by assuming that the NP contribution to the decay $b\rightarrow c\,\mu\,\bar{\nu}$ is identical to the NP contribution to the decay $b\rightarrow c\,e\,\bar{\nu}$. Then experimental constraints on $R_D$/$R_{D^*}$ require the NP WCs in $b\rightarrow c\,\mu\,\bar{\nu}$ transition to be similar in magnitude to those listed in table~\ref{tab4} but of opposite sign.

\section{Conclusions}

In this work we have done a refit of NP expressions for
$R_D$ and $R_{D^*}$ with the new world averages. Since 
these values have  slightly less tension with SM more NP solutions
are allowed. About a third of these solutions, especially those with scalar/pseudoscalar NP operators, do not satisfy the 
constraint from $Br(B_c \to \tau \bar{\nu})$ and hence are rejected. Among the allowed solutions, a number of them are degenarate to one another because they have the same magnitudes of vector and axial-vector couplings in $b\rightarrow c\tau\bar{\nu}$ transition. It is impossible to distinguish between two solutions, which differ from each other only by a sign of the amplitude, by studying only those processes driven by $b\rightarrow c\tau\bar{\nu}$ transition.

 All these NP solutions are still allowed when $R_{J/\psi}$ is included in the fit. Except for the tensor NP solution, they all have $R_{J/\psi} = 0.35-0.36$, significantly smaller than the present central value $0.71$. 
Since the experimental uncertainties are large these predictions do fall within the $90\%$ C.L. range.
However, the following observation is in order. The phase space ratio for $b\rightarrow c\,\tau\,\bar{\nu}:b\rightarrow c\,\mu\,\bar{\nu}$ is $0.37$. If there were no hadronization effects, LFU predicts that $R_D$, $R_{D^*}$ and $R_{J/\psi}$ should all be equal to this ratio. The predicted values of these quantities in SM are indeed different because of different hadronization dynamics. Given that the measured values of $R_D$ and $R_{D^*}$ are larger only by about $20-30\%$ compared to SM, we expect the NP to change the amplitudes by about $10-15\%$. The present central value of $R_{J/\psi}$ is about $2.5$ times the SM prediction. An NP amplitude consistent with $R_D$ and $R_{D^*}$ can change $R_{J/\psi}$ by a maximum of $30\%$. It is impossible to obtain a $100\%$ increase in the value of $R_{J/\psi}$ without a violent disagreement with $R_D$ and/or $R_{D^*}$. Hence, we believe that a future measurement of $R_{J/\psi}$ must necessarily have a smaller central value. If later measurements of $R_{J/\psi}$ find it to be smaller than the SM prediction, then tensor NP is the likely solution.

By performing fit including all available data on $b\rightarrow c\,\tau\,\bar{\nu}$ transition, we identify  all allowed NP solutions and show that there are essentially only four NP solutions. We have also done the calculation using the measurements of $R_D$, $R_{D^*}$, $P_{\tau}$ and $R_{J/\psi}$ for each individual experiment. We note that the need for NP is driven by those measurements where the $\tau$ lepton was not studied. Further, we demonstrate that NP  only in $b\rightarrow c\,\mu\,\bar{\nu}$ {\emph{does not}} provide a viable solution to the $R_D$/$R_{D^*}$ anomaly.

 We also note from table~\ref{tab4} that tensor NP can also be distinguished by means of tau polarization $P_{\tau}$. In ref.~\cite{Alok:2016qyh}, it was shown that the $D^*$ polarization fraction is also effective in distinguishing the tensor NP solution. To make a distinction between the rest of the solutions we need other angular variables such as forward-backward asymmetry and longitudinal-transverse asymmetry. This problem is studied in ref.~\cite{Alok:2018uft}.

\section{Acknowledgement}
SUS thanks the theory group of CERN for their hospitality
when this paper is being finalized. He also thanks Concezio Bozzi
and Greg Ciezarek for valuable discussions.


\begin{thebibliography}{99}

\bibitem{Lees:2012xj} 
  J.~P.~Lees {\it et al.} [BaBar Collaboration],
  Phys.\ Rev.\ Lett.\  {\bf 109}, 101802 (2012)
  [arXiv:1205.5442 [hep-ex]].
  
  \bibitem{Lees:2013uzd} 
  J.~P.~Lees {\it et al.} [BaBar Collaboration],
  Phys.\ Rev.\ D {\bf 88}, no. 7, 072012 (2013)
  [arXiv:1303.0571 [hep-ex]].
  
\bibitem{Huschle:2015rga} 
  M.~Huschle {\it et al.} [Belle Collaboration],
  Phys.\ Rev.\ D {\bf 92}, no. 7, 072014 (2015)
  [arXiv:1507.03233 [hep-ex]].
  
\bibitem{Sato:2016svk} 
  Y.~Sato {\it et al.} [Belle Collaboration],
  Phys.\ Rev.\ D {\bf 94}, no. 7, 072007 (2016)
  [arXiv:1607.07923 [hep-ex]].
  
  \bibitem{Hirose:2016wfn} 
  S.~Hirose {\it et al.} [Belle Collaboration],
  Phys.\ Rev.\ Lett.\  {\bf 118}, no. 21, 211801 (2017)
  [arXiv:1612.00529 [hep-ex]].

\bibitem{Aaij:2015yra} 
  R.~Aaij {\it et al.} [LHCb Collaboration],
  Phys.\ Rev.\ Lett.\  {\bf 115}, no. 11, 111803 (2015)
  [Phys.\ Rev.\ Lett.\  {\bf 115}, no. 15, 159901 (2015)]
  [arXiv:1506.08614 [hep-ex]].

  \bibitem{Aaij:2017uff} 
  R.~Aaij {\it et al.} [LHCb Collaboration],
  arXiv:1708.08856 [hep-ex].
  
  \bibitem{Aaij:2017deq}
  R.~Aaij {\it et al.} [LHCb Collaboration],
  Phys.\ Rev.\ D {\bf 97} (2018) no.7,  072013
  doi:10.1103/PhysRevD.97.072013
  [arXiv:1711.02505 [hep-ex]].
  
\bibitem{Aoki:2016frl} 
  S.~Aoki {\it et al.},
  Eur.\ Phys.\ J.\ C {\bf 77}, no. 2, 112 (2017)
  [arXiv:1607.00299 [hep-lat]].
  

  
  \bibitem{Fajfer:2012vx} 
  S.~Fajfer, J.~F.~Kamenik and I.~Nisandzic,
  Phys.\ Rev.\ D {\bf 85}, 094025 (2012)
  [arXiv:1203.2654 [hep-ph]].
  
  \bibitem{average}
  http://www.slac.stanford.edu/xorg/hfag/semi/fpcp17/RDRDs.html
  

 

  
   \bibitem{Bigi:2016mdz}
  D.~Bigi and P.~Gambino,
  Phys.\ Rev.\ D {\bf 94} (2016) no.9,  094008
  [arXiv:1606.08030 [hep-ph]].
  
  \bibitem{Bernlochner:2017jka}
  F.~U.~Bernlochner, Z.~Ligeti, M.~Papucci and D.~J.~Robinson,
  Phys.\ Rev.\ D {\bf 95} (2017) no.11,  115008
   Erratum: [Phys.\ Rev.\ D {\bf 97} (2018) no.5,  059902]
  [arXiv:1703.05330 [hep-ph]].

 \bibitem{Bigi:2017jbd}
  D.~Bigi, P.~Gambino and S.~Schacht,
  JHEP {\bf 1711} (2017) 061
  [arXiv:1707.09509 [hep-ph]].
  
  \bibitem{Jaiswal:2017rve}
  S.~Jaiswal, S.~Nandi and S.~K.~Patra,
  JHEP {\bf 1712} (2017) 060
  [arXiv:1707.09977 [hep-ph]].
  
  \bibitem{deBoer:2018ipi} 
  S.~de Boer, T.~Kitahara and I.~Nisandzic,
  arXiv:1803.05881 [hep-ph].

  \bibitem{Tanaka:2012nw} 
  M.~Tanaka and R.~Watanabe,
  Phys.\ Rev.\ D {\bf 87}, no. 3, 034028 (2013)
  [arXiv:1212.1878 [hep-ph]].
  

 \bibitem{lhcb-new} 
  LHCb-PAPER-2017-035.
  
  


  
  \bibitem{Dutta:2017xmj} 
  R.~Dutta and A.~Bhol,
  Phys.\ Rev.\ D {\bf 96}, no. 7, 076001 (2017)
  [arXiv:1701.08598 [hep-ph]].
  
  
  \bibitem{Fajfer:2012jt} 
  S.~Fajfer, J.~F.~Kamenik, I.~Nisandzic and J.~Zupan,
  Phys.\ Rev.\ Lett.\  {\bf 109}, 161801 (2012)
  [arXiv:1206.1872 [hep-ph]].
  
    \bibitem{Alonso:2015sja} 
  R.~Alonso, B.~Grinstein and J.~Martin Camalich,
  JHEP {\bf 1510}, 184 (2015)
  [arXiv:1505.05164 [hep-ph]].
  
    
  \bibitem{Ivanov:2017mrj} 
  M.~A.~Ivanov, J.~G.~K\"{o}rner and C.~T.~Tran,
  Phys.\ Rev.\ D {\bf 95}, no. 3, 036021 (2017)
  [arXiv:1701.02937 [hep-ph]].
  
  
   \bibitem{Datta:2012qk} 
  A.~Datta, M.~Duraisamy and D.~Ghosh,
  Phys.\ Rev.\ D {\bf 86}, 034027 (2012)
  [arXiv:1206.3760 [hep-ph]].
  
  \bibitem{Biancofiore:2013ki} 
  P.~Biancofiore, P.~Colangelo and F.~De Fazio,
  Phys.\ Rev.\ D {\bf 87}, no. 7, 074010 (2013)
  [arXiv:1302.1042 [hep-ph]].

\bibitem{Duraisamy:2013kcw} 
  M.~Duraisamy and A.~Datta,
  JHEP {\bf 1309}, 059 (2013)
  [arXiv:1302.7031 [hep-ph]].

  \bibitem{Duraisamy:2014sna} 
  M.~Duraisamy, P.~Sharma and A.~Datta,
  Phys.\ Rev.\ D {\bf 90}, no. 7, 074013 (2014)
  [arXiv:1405.3719 [hep-ph]].

\bibitem{Sakaki:2014sea} 
  Y.~Sakaki, M.~Tanaka, A.~Tayduganov and R.~Watanabe,
  Phys.\ Rev.\ D {\bf 91}, no. 11, 114028 (2015)
  [arXiv:1412.3761 [hep-ph]].
  

 

  \bibitem{Freytsis:2015qca} 
  M.~Freytsis, Z.~Ligeti and J.~T.~Ruderman,
  Phys.\ Rev.\ D {\bf 92}, no. 5, 054018 (2015)
  [arXiv:1506.08896 [hep-ph]].

  \bibitem{Becirevic:2016hea}
  D.~Becirevic, S.~Fajfer, I.~Nisandzic and A.~Tayduganov,
  arXiv:1602.03030 [hep-ph].


\bibitem{Alonso:2016gym} 
  R.~Alonso, A.~Kobach and J.~Martin Camalich,
  Phys.\ Rev.\ D {\bf 94}, no. 9, 094021 (2016)
  [arXiv:1602.07671 [hep-ph]].
  
      \bibitem{Alok:2016qyh} 
  A.~K.~Alok, D.~Kumar, S.~Kumbhakar and S.~U.~Sankar,
  Phys.\ Rev.\ D {\bf 95}, no. 11, 115038 (2017)
  [arXiv:1606.03164 [hep-ph]].
  
    
  \bibitem{Ivanov:2016qtw} 
  M.~A.~Ivanov, J.~G.~K\"{o}rner and C.~T.~Tran,
  Phys.\ Rev.\ D {\bf 94}, no. 9, 094028 (2016)
  [arXiv:1607.02932 [hep-ph]].
  
    \bibitem{Ligeti:2016npd} 
  Z.~Ligeti, M.~Papucci and D.~J.~Robinson,
  JHEP {\bf 1701}, 083 (2017)
  [arXiv:1610.02045 [hep-ph]].
  
  \bibitem{Bardhan:2016uhr} 
  D.~Bardhan, P.~Byakti and D.~Ghosh,
  JHEP {\bf 1701}, 125 (2017)
  [arXiv:1610.03038 [hep-ph]].
  
\bibitem{Kim:2016yth} 
  C.~S.~Kim, G.~Lopez-Castro, S.~L.~Tostado and A.~Vicente,
  Phys.\ Rev.\ D {\bf 95}, no. 1, 013003 (2017)
  [arXiv:1610.04190 [hep-ph]].
  
 
  \bibitem{Dutta:2016eml} 
  R.~Dutta and A.~Bhol,
  Phys.\ Rev.\ D {\bf 96}, no. 3, 036012 (2017)
  [arXiv:1611.00231 [hep-ph]].
  
  \bibitem{Bhattacharya:2016zcw} 
  S.~Bhattacharya, S.~Nandi and S.~K.~Patra,
  Phys.\ Rev.\ D {\bf 95}, no. 7, 075012 (2017)
  [arXiv:1611.04605 [hep-ph]].
  
    \bibitem{Alonso:2016oyd} 
  R.~Alonso, B.~Grinstein and J.~Martin Camalich,
  Phys.\ Rev.\ Lett.\  {\bf 118}, no. 8, 081802 (2017)
  [arXiv:1611.06676 [hep-ph]].

    
  \bibitem{Alonso:2017ktd} 
  R.~Alonso, J.~Martin Camalich and S.~Westhoff,
  Phys.\ Rev.\ D {\bf 95}, no. 9, 093006 (2017)
  [arXiv:1702.02773 [hep-ph]].
  
  \bibitem{Jung:2018lfu} 
  M.~Jung and D.~M.~Straub,
  arXiv:1801.01112 [hep-ph].
  
  \bibitem{Colangelo:2018cnj} 
  P.~Colangelo and F.~De Fazio,
  arXiv:1801.10468 [hep-ph].





  
      \bibitem{Sakaki:2013bfa} 
  Y.~Sakaki, M.~Tanaka, A.~Tayduganov and R.~Watanabe,
  Phys.\ Rev.\ D {\bf 88}, no. 9, 094012 (2013)
  [arXiv:1309.0301 [hep-ph]].
  
    \bibitem{Fajfer:2015ycq} 
  S.~Fajfer and N.~Ko\v{s}nik,
  Phys.\ Lett.\ B {\bf 755}, 270 (2016)
  [arXiv:1511.06024 [hep-ph]].
  

\bibitem{Bauer:2015knc} 
  M.~Bauer and M.~Neubert,
  Phys.\ Rev.\ Lett.\  {\bf 116}, no. 14, 141802 (2016)
  [arXiv:1511.01900 [hep-ph]].
  
  
  \bibitem{Barbieri:2015yvd} 
  R.~Barbieri, G.~Isidori, A.~Pattori and F.~Senia,
  Eur.\ Phys.\ J.\ C {\bf 76}, no. 2, 67 (2016)
  [arXiv:1512.01560 [hep-ph]].
 
   
      \bibitem{Dorsner:2016wpm} 
  I.~Dor\v{s}ner, S.~Fajfer, A.~Greljo, J.~F.~Kamenik and N.~Ko\v{s}nik,
  Phys.\ Rept.\  {\bf 641}, 1 (2016)
  [arXiv:1603.04993 [hep-ph]].
  
  
      \bibitem{Li:2016vvp} 
  X.~Q.~Li, Y.~D.~Yang and X.~Zhang,
  JHEP {\bf 1608}, 054 (2016)
  [arXiv:1605.09308 [hep-ph]].
  
    \bibitem{Sahoo:2016pet} 
  S.~Sahoo, R.~Mohanta and A.~K.~Giri,
  Phys.\ Rev.\ D {\bf 95}, no. 3, 035027 (2017)
  [arXiv:1609.04367 [hep-ph]].
  
  \bibitem{Bhattacharya:2016mcc} 
  B.~Bhattacharya, A.~Datta, J.~P.~Gu\'{e}vin, D.~London and R.~Watanabe,
  JHEP {\bf 1701}, 015 (2017)
  [arXiv:1609.09078 [hep-ph]].
  
    
  \bibitem{Barbieri:2016las} 
  R.~Barbieri, C.~W.~Murphy and F.~Senia,
  Eur.\ Phys.\ J.\ C {\bf 77}, no. 1, 8 (2017)
  [arXiv:1611.04930 [hep-ph]].
 

  \bibitem{Chen:2017hir} 
  C.~H.~Chen, T.~Nomura and H.~Okada,
  arXiv:1703.03251 [hep-ph].
  
    \bibitem{Crivellin:2017zlb} 
  A.~Crivellin, D.~M{\"{u}}ller and T.~Ota,
  JHEP {\bf 1709}, 040 (2017)
  [arXiv:1703.09226 [hep-ph]].
  
    
  \bibitem{Alok:2017jaf} 
  A.~K.~Alok, D.~Kumar, J.~Kumar and R.~Sharma,
  arXiv:1704.07347 [hep-ph].
  
    \bibitem{Calibbi:2017qbu} 
  L.~Calibbi, A.~Crivellin and T.~Li,
  arXiv:1709.00692 [hep-ph].
  
  
    
  \bibitem{Crivellin:2012ye} 
  A.~Crivellin, C.~Greub and A.~Kokulu,
  Phys.\ Rev.\ D {\bf 86}, 054014 (2012)
  [arXiv:1206.2634 [hep-ph]].
  
  \bibitem{Celis:2012dk} 
  A.~Celis, M.~Jung, X.~Q.~Li and A.~Pich,
  JHEP {\bf 1301}, 054 (2013)
  [arXiv:1210.8443 [hep-ph]].
  
  
    \bibitem{Crivellin:2015hha} 
  A.~Crivellin, J.~Heeck and P.~Stoffer,
  Phys.\ Rev.\ Lett.\  {\bf 116}, no. 8, 081801 (2016)
  [arXiv:1507.07567 [hep-ph]].
  
    \bibitem{Wang:2016ggf} 
  L.~Wang, J.~M.~Yang and Y.~Zhang,
  arXiv:1610.05681 [hep-ph].

  
  \bibitem{Celis:2016azn} 
  A.~Celis, M.~Jung, X.~Q.~Li and A.~Pich,
  Phys.\ Lett.\ B {\bf 771}, 168 (2017)
  [arXiv:1612.07757 [hep-ph]].
  
  \bibitem{Ko:2017lzd} 
  P.~Ko, Y.~Omura, Y.~Shigekami and C.~Yu,
  Phys.\ Rev.\ D {\bf 95}, no. 11, 115040 (2017)
  [arXiv:1702.08666 [hep-ph]].
  
    \bibitem{Iguro:2017ysu} 
  S.~Iguro and K.~Tobe,
  arXiv:1708.06176 [hep-ph].

\bibitem{Biswas:2018jun} 
  A.~Biswas, D.~K.~Ghosh, A.~Shaw and S.~K.~Patra,
  arXiv:1801.03375 [hep-ph].
  
  \bibitem{Martinez:2018ynq} 
  R.~Martinez, C.~F.~Sierra and G.~Valencia,
  arXiv:1805.04098 [hep-ph].
  
 \bibitem{Greljo:2015mma} 
  A.~Greljo, G.~Isidori and D.~Marzocca,
  JHEP {\bf 1507}, 142 (2015)
  [arXiv:1506.01705 [hep-ph]].

 
    \bibitem{Boucenna:2016wpr} 
  S.~M.~Boucenna, A.~Celis, J.~Fuentes-Martin, A.~Vicente and J.~Virto,
  Phys.\ Lett.\ B {\bf 760}, 214 (2016)
  [arXiv:1604.03088 [hep-ph]].
  
    \bibitem{Matsuzaki:2017bpp} 
  S.~Matsuzaki, K.~Nishiwaki and R.~Watanabe,
  JHEP {\bf 1708}, 145 (2017)
  [arXiv:1706.01463 [hep-ph]].
  
\bibitem{Asadi:2018wea} 
  P.~Asadi, M.~R.~Buckley and D.~Shih,
  arXiv:1804.04135 [hep-ph].


    \bibitem{Das:2016vkr} 
  D.~Das, C.~Hati, G.~Kumar and N.~Mahajan,
  Phys.\ Rev.\ D {\bf 94}, 055034 (2016)
  [arXiv:1605.06313 [hep-ph]].
  
    \bibitem{Deshpand:2016cpw} 
  N.~G.~Deshpande and X.~G.~He,
  Eur.\ Phys.\ J.\ C {\bf 77}, no. 2, 134 (2017)
  [arXiv:1608.04817 [hep-ph]].
  
  
    \bibitem{Cvetic:2017gkt} 
  G.~Cveti{\v{c}}, F.~Halzen, C.~S.~Kim and S.~Oh,
  arXiv:1702.04335 [hep-ph].
  
    \bibitem{Aloni:2017eny} 
  D.~Aloni, A.~Efrati, Y.~Grossman and Y.~Nir,
  JHEP {\bf 1706}, 019 (2017)
  [arXiv:1702.07356 [hep-ph]].
  
  \bibitem{Megias:2017ove} 
  E.~Megias, M.~Quiros and L.~Salas,
  JHEP {\bf 1707}, 102 (2017)
  [arXiv:1703.06019 [hep-ph]].

  
  \bibitem{Altmannshofer:2017poe} 
  W.~Altmannshofer, P.~S.~B.~Dev and A.~Soni,
  arXiv:1704.06659 [hep-ph].
  
    \bibitem{Feruglio:2017rjo} 
  F.~Feruglio, P.~Paradisi and A.~Pattori,
  JHEP {\bf 1709}, 061 (2017)
  [arXiv:1705.00929 [hep-ph]].
  
    \bibitem{Choudhury:2017qyt} 
  D.~Choudhury, A.~Kundu, R.~Mandal and R.~Sinha,
  arXiv:1706.08437 [hep-ph].
  
  \bibitem{Cline:2017ihf} 
  J.~M.~Cline and J.~Martin Camalich,
  Phys.\ Rev.\ D {\bf 96}, no. 5, 055036 (2017)
  [arXiv:1706.08510 [hep-ph]].
  
  \bibitem{Watanabe:2017mip} 
  R.~Watanabe,
  arXiv:1709.08644 [hep-ph].
  
  \bibitem{Chauhan:2017uil} 
  B.~Chauhan and B.~Kindra,
  arXiv:1709.09989 [hep-ph].
  
  \bibitem{Dutta:2017wpq} 
  R.~Dutta,
  arXiv:1710.00351 [hep-ph].
  
  \bibitem{Tran:2018kuv} 
  C.~T.~Tran, M.~A.~Ivanov, J.~G.~K{\"o}rner and P.~Santorelli,
  Phys.\ Rev.\ D {\bf 97}, no. 5, 054014 (2018)
  [arXiv:1801.06927 [hep-ph]].
  
  \bibitem{Wei:2018vmk} 
  B.~Wei, J.~Zhu, J.~H.~Shen, R.~M.~Wang and G.~R.~Lu,
  arXiv:1801.00917 [hep-ph].
  
  \bibitem{Rui:2018kqr} 
  Z.~Rui, J.~Zhang and L.~L.~Zhang,
  arXiv:1806.00796 [hep-ph].
 
\bibitem{Greljo:2018ogz}
  A.~Greljo, D.~J.~Robinson, B.~Shakya and J.~Zupan,
  arXiv:1804.04642 [hep-ph].
  
\bibitem{Robinson:2018gza}
  D.~Robinson, B.~Shakya and J.~Zupan,
  arXiv:1807.04753 [hep-ph].
  
  
  
  \bibitem{Davidson:1993qk}
  S.~Davidson, D.~C.~Bailey and B.~A.~Campbell,
  Z.\ Phys.\ C {\bf 61} (1994) 613
  doi:10.1007/BF01552629
  [hep-ph/9309310].
  
  \bibitem{Davidson:2010uu}
  S.~Davidson and S.~Descotes-Genon,
  JHEP {\bf 1011} (2010) 073
  doi:10.1007/JHEP11(2010)073
  [arXiv:1009.1998 [hep-ph]].
 
  
\bibitem{Caprini:1997mu}
  I.~Caprini, L.~Lellouch and M.~Neubert,
  Nucl.\ Phys.\ B {\bf 530} (1998) 153
  [hep-ph/9712417].
  
  \bibitem{Bailey:2014tva}
  J.~A.~Bailey {\it et al.} [Fermilab Lattice and MILC Collaborations],
  Phys.\ Rev.\ D {\bf 89} (2014) no.11,  114504
  [arXiv:1403.0635 [hep-lat]].
  
  \bibitem{Amhis:2016xyh}
  Y.~Amhis {\it et al.} [HFLAV Collaboration],
  Eur.\ Phys.\ J.\ C {\bf 77} (2017) no.12,  895
  [arXiv:1612.07233 [hep-ex]].
  
  
    \bibitem{minuit1}
	F. James and M. Roos, ''Minuit: A System for Function Minimization and 		    Analysis of the
    Parameter Errors and Correlations,'' Comput. Phys. Commun. 10, 343     (1975). 

    \bibitem{minuit2}
    F. James, ''MINUIT Function Minimization and Error Analysis: Reference Manual Version
94.1,'' CERN-D-506, CERN-D506.

  


  \bibitem{Wen-Fei:2013uea} 
  W.~F.~Wang, Y.~Y.~Fan and Z.~J.~Xiao,
  Chin.\ Phys.\ C {\bf 37}, 093102 (2013)
  [arXiv:1212.5903 [hep-ph]].
  
    \bibitem{Colquhoun:2015oha}
  B.~Colquhoun {\it et al.} [HPQCD Collaboration],
  Phys.\ Rev.\ D {\bf 91} (2015) no.11,  114509
  [arXiv:1503.05762 [hep-lat]].
  
  \bibitem{pdg}
  C. Patrignani  {\it et al.} [Particle Data Group],
  Chin.\ Phys.\ C {\bf 40}, 100001 (2016).
 
  
  

  
  \bibitem{Akeroyd:2017mhr} 
  A.~G.~Akeroyd and C.~H.~Chen,
  Phys.\ Rev.\ D {\bf 96}, no. 7, 075011 (2017)
  [arXiv:1708.04072 [hep-ph]].
  


  
  \bibitem{Glattauer:2015teq} 
  R.~Glattauer {\it et al.} [Belle Collaboration],
  Phys.\ Rev.\ D {\bf 93}, no. 3, 032006 (2016)
  [arXiv:1510.03657 [hep-ex]].
  
  \bibitem{Abdesselam:2017kjf} 
  A.~Abdesselam {\it et al.} [Belle Collaboration],
  arXiv:1702.01521 [hep-ex].
  
\bibitem{Alok:2018uft}
  A.~K.~Alok, D.~Kumar, S.~Kumbhakar and S.~Uma Sankar,
  Phys.\ Lett.\ B {\bf 784} (2018) 16
  [arXiv:1804.08078 [hep-ph]].


\end{thebibliography}
\end{document}